\documentclass[a4paper]{jpconf}
\usepackage{graphicx}
\begin{document}
\title{Synthetic Paths to the Heaviest Elements}

\author{W. Loveland}

\address{Oregon State University, Corvallis, OR 97331 USA}

\ead{lovelanw@onid.orst.edu}

\begin{abstract}
The cross section for producing a heavy reaction product, $\sigma$$_{EVR}$, can be represented by the equation
\begin{displaymath}
\sigma _{EVR}=\sum_{J=0}^{J_{\max }}\sigma
_{capture}(E_{c.m.},J)P_{CN}(E^{*},J) W_{sur}(E^{*},J)
\end{displaymath}
where $\sigma _{capture}(E_{c.m.},J)$ is the capture cross section at center of mass energy E$_{c.m.}$ and spin J. P$_{CN}$ is the probability that the projectile-target system will evolve from the contact configuration  inside the
fission saddle point to form a completely fused system rather than
re-separating (quasifission, fast fission). W$_{sur}$ is
the probability that the completely fused system will de-excite by neutron emission rather than fission. 
 I discuss results of experiments that characterize these quantities in heavy element synthesis reactions.  I also discuss the possibilities of synthesizing heavy nuclei using damped collisions and reactions using radioactive beams.

 \end{abstract}

\section{Complete fusion}

\subsection{Overview}

Most heavy element synthesis reactions to date have involved complete fusion reactions. For these reactions, the cross section for producing a heavy evaporation residue, $\sigma$$_{EVR}$, can be written as
\setcounter{equation}{0}
\begin{equation}
\sigma _{EVR}=\sum_{J=0}^{J_{\max }}\sigma
_{capture}(E_{c.m.},J)P_{CN}(E^{*},J) W_{sur}(E^{*},J)
\end{equation}
where $\sigma _{capture}(E_{c.m.},J)$ is the capture cross section at center of mass energy E$_{c.m.}$ and spin J. P$_{CN}$ is the probability that the projectile-target system will evolve from the contact configuration  inside the
fission saddle point to form a completely fused system rather than
re-separating (quasifission, fast fission). W$_{sur}$ is
the probability that the completely fused system will de-excite by neutron emission rather than fission. For a quantitative understanding of the synthesis of new heavy nuclei, one needs to understand the spin and isospin dependence of $\sigma _{capture}$, P$_{CN}$, and W$_{sur}$ for the reaction system under study.

The first question we might ask ourselves about heavy element fusion cross sections is how well can we predict them. The answer to this question is ``very well".  There are a number of predictions of the evaporation residue cross sections in hot and cold fusion reactions that agree remarkably well with measured evaporation residue cross sections [1]. This is a significant achievement considering that the measured cross sections span at least six orders of magnitude.  However, taking cold fusion reactions as an example, we show, in Figure 1, a typical set of predictions of the evaporation residue cross sections in cold fusion reactions [2-7].  The measured data are well described by all the predictions.  However if we look at the values (Fig. 2) of the quantities in equation [1] such as P$_{CN}$ we see that the values of P$_{CN}$  can differ by orders of magnitude in the various calculations even though they all give the same (correct) answer for the value of the evaporation residue cross section..  We are engaged in efforts to measure the individual factors in equation (2) to better constrain/advise models to calculate heavy element evaporation residue cross sections.

\begin{figure}[h]
\begin{minipage}{18pc}
\includegraphics[width=18pc]{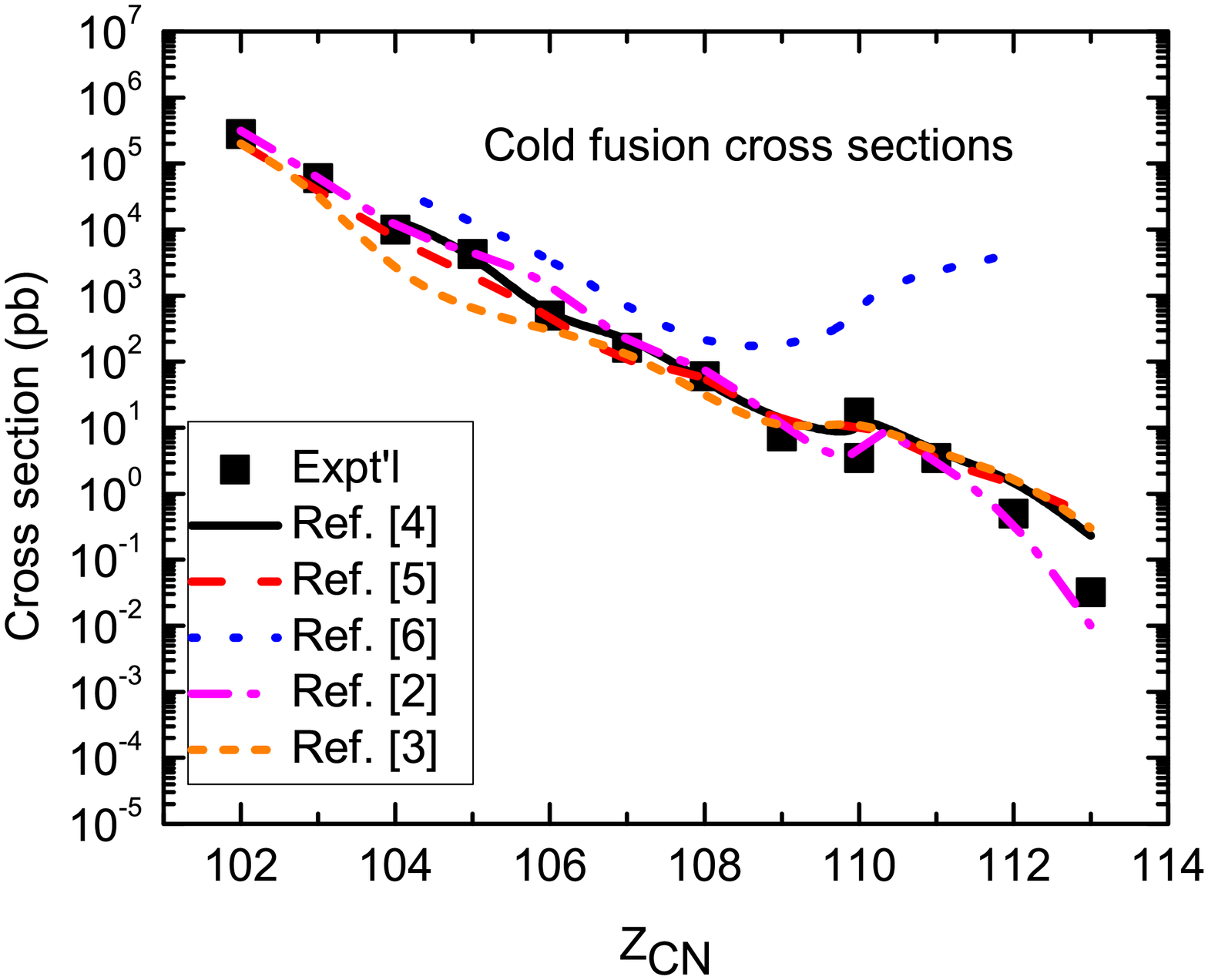}
\caption{\label{fig 1a}Typical predictions of the formation cross sections of elements 102-113 using cold fusion reactions.}
\end{minipage}\hspace{2pc}%
\begin{minipage}{14pc}
\includegraphics[width=18pc]{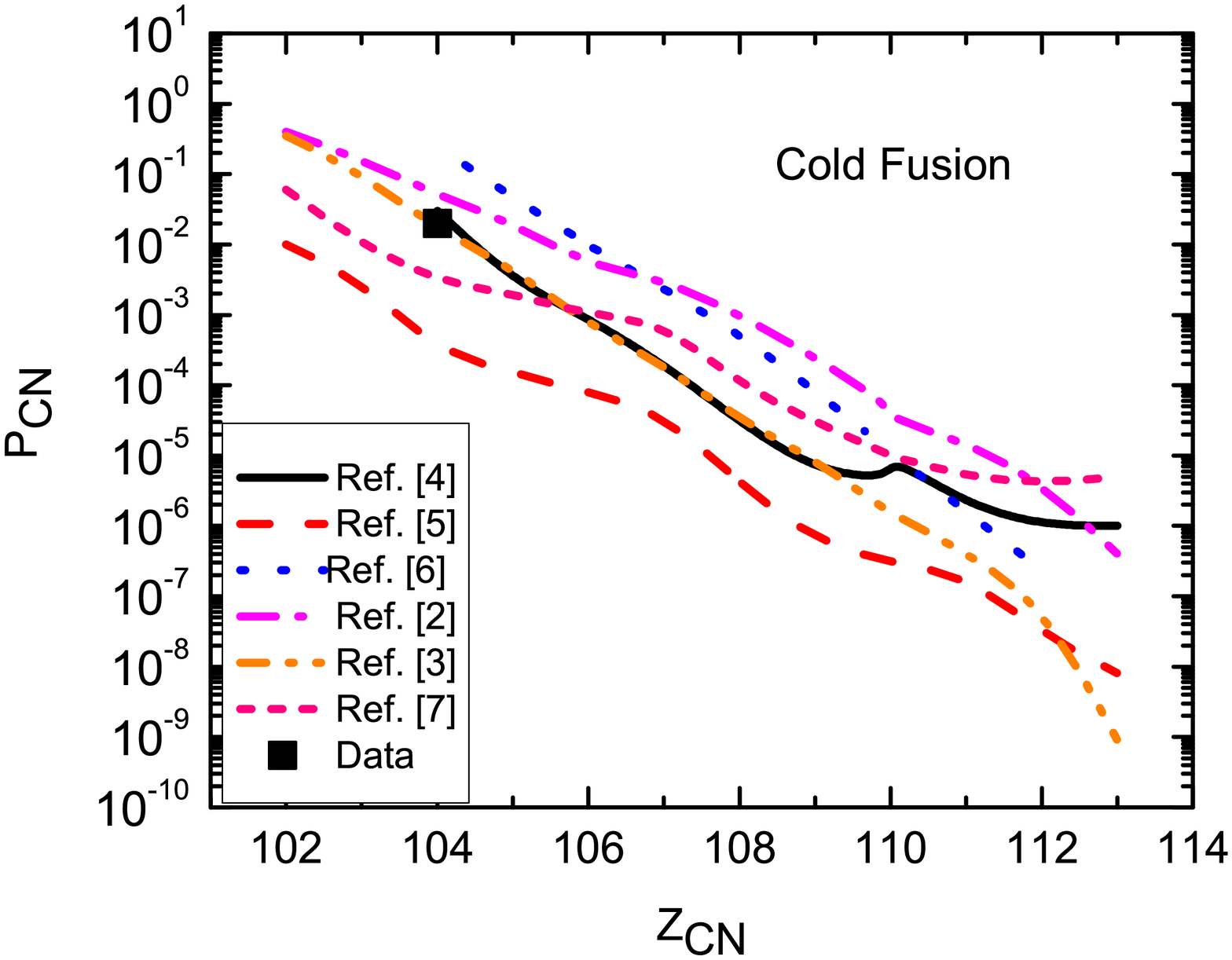}
\caption{\label{fig 1b}Comparison of predictions of P$_{CN}$ for these cold fusion reactions}
\end{minipage} 
\end{figure}
\begin{figure}[h]
\begin{minipage}{40pc}
\begin{center}
\includegraphics[width=40pc]{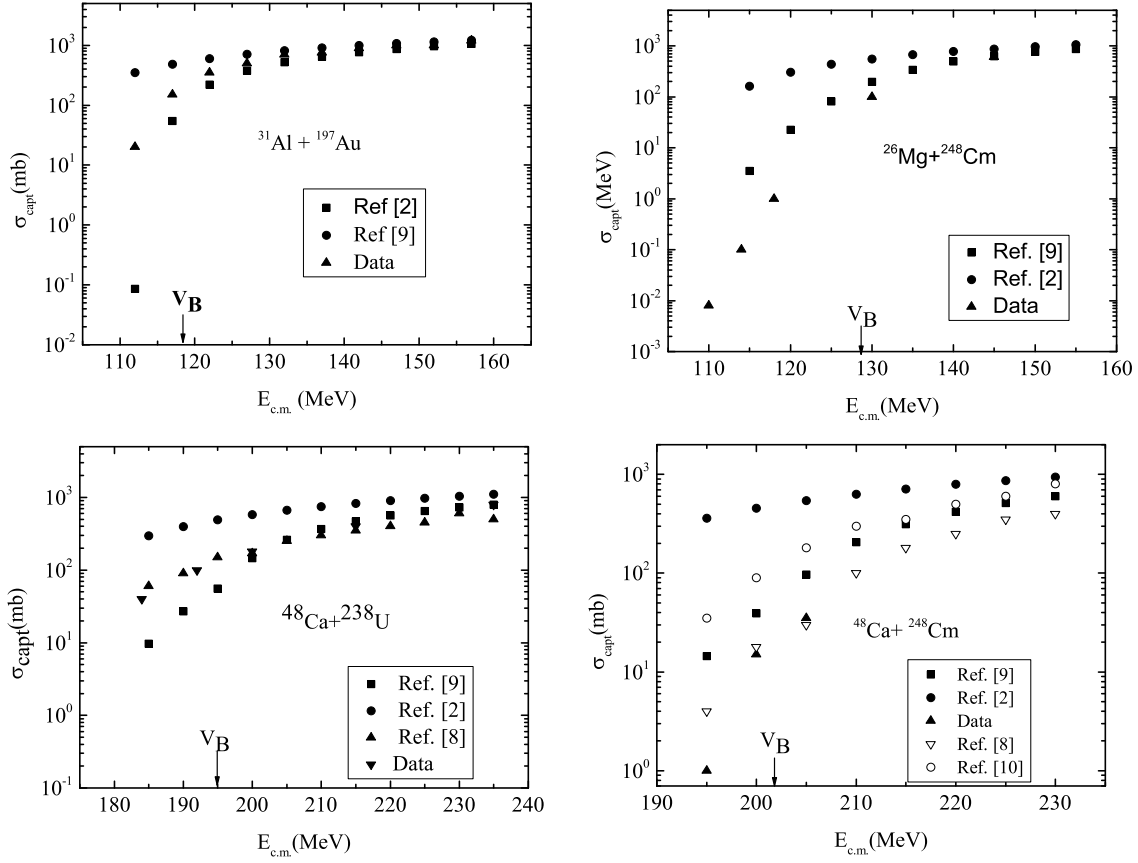}
\end{center}
\caption{\label{fig3}Sample predictions of capture cross sections for reactions synthesizing heavy elements.}
\end{minipage} 
\end{figure}

\subsection{Capture cross sections}
The capture cross section is, in the language of coupled channel calculations, the ``barrier crossing" cross section.  It is the sum of the quasifission, fast fission, fusion-fission and fusion-evaporation residue cross sections.  The latter cross section is so small for the systems studied in this work that it is neglected.  The barriers involved are the interaction barriers and not the fusion barriers.  There are several models for capture cross sections [2,8-11].  Each of them has been tested against a number of measurements of capture cross sections for reactions that, mostly, do not lead to the formation of the heaviest nuclei.  In general, these models are able to predict the magnitudes of the capture cross sections with 50$\%$ and the values of the interaction barriers within 20$\%$.  However, when one compares the predictions of these models with measured data for capture cross sections leading to the formation of heavy elements (Fig 3) the agreement between prediction and data is, at best, a factor of two.  In detail, the calculated and measured cross sections are in reasonable agreement above the barrier, except for the formation of the heaviest elements but significant differences are observed for cross sections near or below the barrier.  This conclusion is similar to that reached in [12] where the sub-barrier capture cross sections were estimated to be uncertain to an order of magnitude with much better agreement between prediction and data above the barrier.  Amongst the many methods for estimating capture cross sections, the use of coupled channels calculations [9,20] appears to do the best job of predicting the capture cross sections.

\subsection{Survival probabilities}
The calculation of the survival probabilities involves a well understood formalism in which the principal uncertainty is the height of the assumed fission barriers.  In the best local approximation of these barrier heights [13], the average uncertainty in the fission barrier heights for the heaviest nuclei was 0.4 MeV, with a maximum deviation between estimated and deduced barrier heights being 1.0 MeV.

The survival probability W$_{sur}$ can be written as 
\begin{equation}
W_{sur}=P_{xn}(E_{CN}^{\ast })\prod\limits_{i=1}^{i_{\max }=x}\left( \frac{%
\Gamma _{n}}{\Gamma _{n}+\Gamma _{f}}\right) _{i,E^{\ast }}
\end{equation}
where the index i is equal to the number of emitted neutrons and P$_{xn}$ is
the probability of emitting exactly x neutrons [14]. In evaluating the excitation energy in  equation [2], we start at the excitation energy E* of the completely fused system and reduce it for each evaporation step by the binding energy of the emitted neutron and an assumed neutron kinetic energy of 2T where T (=(E*/a)$^{1/2}$) is the temperature of the emitting system.   For
calculating $\Gamma _{n}/\Gamma _{f}$, we have used the classical formalism
from Vandenbosch and Huizenga [15]
\begin{equation}
\frac{\Gamma _{n}}{\Gamma _{f}}=\frac{4A^{2/3}\left( E^{\ast }-B_{n}\right) 
}{k\left[ 2a^{1/2}\left( E^{\ast }-B_{f}\right) ^{1/2}-1\right] }\exp \left[
2a^{1/2}\left( E^{\ast }-B_{n}\right) ^{1/2}-2a^{1/2}\left( E^{\ast
}-B_{f}\right) ^{1/2}\right] 
\end{equation}
The constants k and a are taken to \ be 9.8 MeV and (A/12) MeV$^{-1}$, respectively. \ The
fission barriers B$_{f}$ are written as the sum of liquid drop, B$_{f}^{LD}$,
and shell correction terms as
\begin{equation}
B_{f}(E_{CN}^{\ast })=B_{f}^{LD}+U_{shell}
\end{equation}
where the shell correction energies , U$_{shell}$, to the LDM barriers are
taken from [16] , and the liquid drop barriers are taken from [17].  Neutron binding energies, B$_{n}$ are taken from [16]. The fade-out of the shell corrections with increasing excitation energy is treated through the level density parameter using the method of  Ignatyuk et al. [18] as
\begin{equation}
a=\widetilde{a}\left[ 1+\delta E\frac{1-\exp (-\gamma E)}{E}\right] 
\end{equation}
\begin{equation}
\widetilde{a}=0.073A+0.095B_{s}(\beta _{2})A^{2/3}
\end {equation}
where the shell damping parameter is taken to be 0.061.  Variation of this parameter from system to system has been demonstrated by recent calculations [19] but, lacking firm guidance of how to specify the variation of this parameter, I have chosen to keep it constant.
Collective enhancement effects of the level density are important for both deformed and spherical nuclei as are their dependence on excitation energy. [21,22].  I use the formalism of ref. [9] to express these effects via the equations
\begin{equation}
K_{coll}=K_{rot}(E)\varphi (\beta _{2})+E_{vib}(E)\cdot (1-\varphi(\beta _{2}))
\end{equation}
\begin{equation}
\varphi (\beta _{2})=\left[ 1+\exp \left( \frac{\beta _{2}^{0}-\left\vert\beta _{2}\right\vert }{\Delta \beta _{2}}\right) \right] ^{-1}
\end{equation}
\begin{equation}
K_{rot(vib)}(E)=\frac{K_{rot(vib)}-1}{1+\left[ \left( E-E_{\alpha }\right)/\Delta E_{\alpha }\right] }+1
\end{equation}
\begin{equation}
K_{rot}=\frac{J_{\bot }T}{\hbar ^{2}}
\end{equation}
\begin{equation}
K_{vib}=\exp (0.0555A^{2/3}T^{4/3})
\end{equation}

I  took a group of about 75 reactions where there are well measured evaporation residue cross sections for the production of elements 98-108 [23]  and where the product Z$_{1}$Z$_{2}$ was $\le$ 1000 (to insure P$_{CN}$ =1).  I calculated $\sigma$$_{capture}$ using coupled channels calculations and calculated W$_{sur}$ using the formalism outlined above leading to a calculated value of the evaporation residue cross section.  The ratio of the measured to calculated evaporation residue cross sections was 6.5.  If I assume that the capture cross sections are uncertain to a factor of 2, then the survival probabilities are uncertain to a factor of 3-4.  

\subsection{Fusion probabilities}
The fusion probability, P$_{CN}$, is the most poorly known quantity in equation [1].  It is difficult to measure and there is considerable uncertainty (see Figure 2) about how to calculate this quantity.  The essence of measuring P$_{CN}$ is to determine the fusion-fission cross section in the presence of the quasi-fission and fast fission cross sections.  

The separation of quasifission from complete fusion-fission can be done in a variety of ways.  The first method involves measuring the width of the fission mass distributions [24-26] .  One assumes that fusion-fission gives symmetric mass distributions while quasifission gives rise to very asymmetric mass distributions.  This distinction is roughly true but can be problematic in some situations where there is evidence of quasifission resulting in symmetric mass distributions [27,28].

A second method is based upon measuring the fission fragment angular distributions [29,30].  Fusion-fission is assumed to be described by the ordinary transition state model of fission angular distributions while quasifission is described by a somewhat arbitrary strongly fore-aft peaked distribution.  This method appears to be very reliable, although some controversy exists over the possibility that, in some reactions, the K distribution at the saddle point is not fully equilibrated.

A third method, suggested by the group at Australian National University (ANU) group and others [31-33], involves defining a reduced cross section, $\widetilde{\sigma }$, as $\sigma$$_{EVR}$/$\pi$($\lambda$/2$\pi$)$^{2}$.  One then measures the evaporation residue yields in a series of reactions of differing asymmetry that produce the same compound nucleus.  Each reaction is run at a high excitation energy (E$^{*}$ $\geq$ 40 MeV) where P$_{CN}$ is roughly independent of excitation energy (see below).  One assumes that  all relevant partial waves are ``saturated".  One invokes the Bohr independence hypothesis that the decay of the completely fused system (W$_{sur}$) is independent of its mode of formation.  Then, roughly speaking, the ratio of the reduced cross sections, $\widetilde{\sigma }$, is the ratio of the P$_{CN}$ factors for each reaction.  If one of the reacting systems is very asymmetric (P$_{CN}$ =1), then one can get absolute values of P$_{CN}$.  One strength of this method is that it is nominally useful for systems where P$_{CN}$ is less than 0.01 and where it is difficult to determine P$_{CN}$ by measuring mass-angle correlations.

Zagrebaev and Greiner [1] have suggested the following functional form for the excitation energy dependence of P$_{CN}$
\begin{equation}
P_{CN}(E^{\ast },J)=\frac{P_{CN}^{0}}{1+\exp \left[ \frac{E_{B}^{\ast
}-E_{int}^{\ast }(J)}{\Delta }\right] }
\end{equation}
where P$_{CN}^{0}$ is the fissility dependent value of P$_{CN}$ at zero excitation energy, E$_{B}^{\ast}$ is the excitation energy at the Bass barrier, E$_{int}^{\ast }$(J) is the internal excitation energy (E$_{c.m.}$+Q-E$_{rot}$(J)) and $\Delta$ is 4 MeV.  Figure 4 shows a comparison of the data of Knyazheva [34] with the predictions of this formula.  The agreement between the predicted and measured values of P$_{CN}$ is excellent.  (There are alternate treatments of these data that give different values of the dependence of P$_{CN}$ upon excitation energy [35].)

\begin{figure}[h]
\begin{minipage}{28pc}
\begin{center}
\includegraphics[width=28pc]{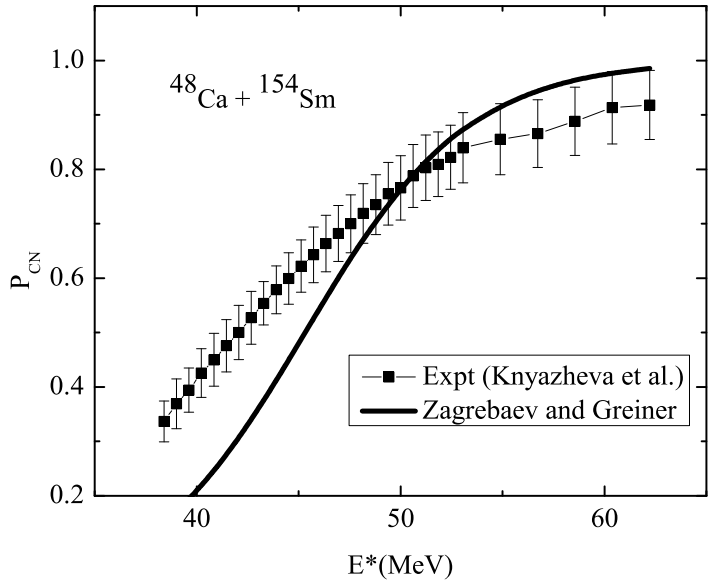}
\end{center}
\caption{\label{fig4}Comparison of measured [34] and predicted [1] values of P$_{CN}$ for the reaction of $^{48}$Ca with $^{154}$Sm.}
\end{minipage} 
\end{figure}

In Figure 5, I show a comparison of the measured values of P$_{CN}$ for the $^{50}$Ti + $^{208}$Pb reaction.  At an excitation energy of about 33 MeV one sees the results of measurement of P$_{CN}$ by the angular distribution method [44] and the mass distribution method [43].  The encouraging approximate agreement between the two measurements is a measure of how well one can measure P$_{CN}$.  One next directs attention to the points at an excitation energy of 15 MeV, the excitation energy used in cold fusion reactions.   The calculated values of P$_{CN}$ differ by two orders of magnitude with a modest number of estimates near the measured value of 0.02.  The estimates of the excitation energy dependence of P$_{CN}$ are in rough agreement with the experimental data.

\begin{figure}[h]
\begin{minipage}{28pc}
\begin{center}
\includegraphics[width=28pc]{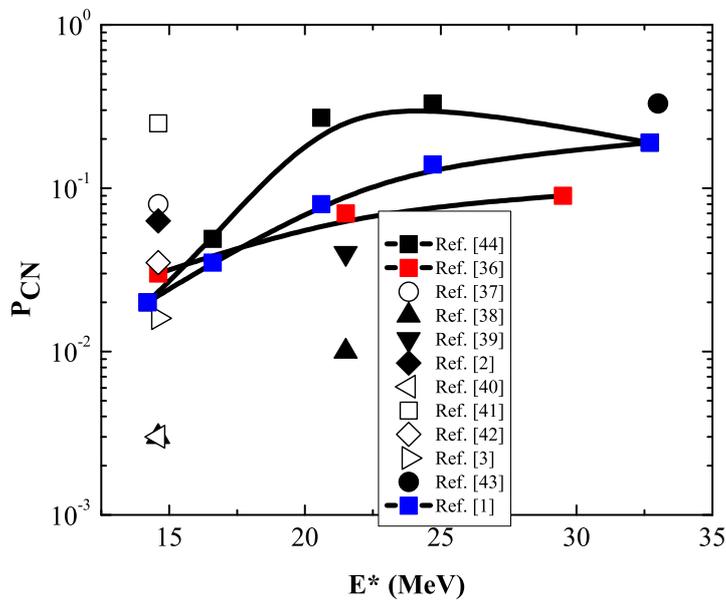}
\end{center}
\caption{\label{fig5}Comparison of measured [44] and predicted values of P$_{CN}$ for the reaction of $^{50}$Ti with $^{208}$Pb}
\end{minipage} 
\end{figure}

To understand the dependence of P$_{CN}$ upon fissility (or some other scaling variable that reflects the entrance channel), I compiled a current list of measured values of P$_{CN}$ using the angular distribution method (AD) the mass distribution method (MY), or the mass-angle correlation (MAD) that is shown in Table 1.  The scaling parameter z equals Z$_{1}$Z$_{2}$/(A$_{1}$$^{1/3}$+A$_{2}$$^{1/3}$).  

\begin{center}
\begin{table}[ht]
\vspace{0.5cm}
\caption{Measurements of P$_{CN}$}
\begin{tabular}{ccccccccccc}
Proj.&Target&CN&E$_{c.m.}$(MeV) &E*(MeV)&Z$_{1}$Z$_{2}$&z&x$_{eff}$&P$_{CN}$& Ref&Method   \\
\hline
$^{11}$B&$^{204}$Pb&$^{215}$At&48-60&31-43&410&50.6&0.325&1-1&[45]&AD \\
$^{16}$O&$^{186}$W&$^{202}$Pb&70-121&48-100&592&72.0&0.420&1-1&[46]&MAD \\
$^{18}$O&$^{197}$Au&$^{215}$At&71-89&39-56&632&74.9&0.413&1-1&[45]&AD \\
$^{19}$F&$^{208}$Pb&$^{227}$Pa&101-174&51-124&738&85.89&0.459&0.78-0.83&[29]&AD \\
$^{24}$Mg&$^{208}$Pb&$^{232}$Pu&126-188&52-114&984&111.7&0.549&0.64-0.71&[29]&AD \\
$^{48}$Ca&$^{144}$Sm&$^{192}$W&141-167&38-64&1240&139.7&0.600&1-1&[46]&MAD \\
$^{28}$Si&$^{208}$Pb&$^{236}$Cm&141-229&50-138&1148&128.1&0.597&0.37-0.63&[29]&AD \\
$^{26}$Mg&$^{248}$Cm&$^{274}$Hs&119-146&37-64&1152&124.6&0.572&0.6&[48]&MY \\
$^{32}$S&$^{182}$W&$^{214}$Th&141-221&56-136&1184&133.9&0.613&0.14-0.51&[49]&AD \\
$^{48}$Ca&$^{154}$Sm&$^{202}$Pb&139-185&49-95&1240&137.9&0.594&0.55-0.94&[46]&MAD \\
$^{40}$Ca&$^{154}$Sm&$^{194}$Pb&139-158&56-75&1240&141.3&0.633&0.89-0.98&[46]&MAD \\
$^{32}$S&$^{197}$Au&$^{229}$Am&151-194&60-98&1264&140.6&0.641&0.42-0.58&[45]&AD\\
$^{32}$S&$^{208}$Pb&$^{240}$Cf&172-217&66-111&1312&144.2&0.641&0.45-0.46&[29]&AD \\
$^{36}$S&$^{238}$U&$^{274}$Hs&153-173&36-56&1472&155.0&0.647&0.043-0.3&[48]&MY \\
$^{50}$Ti&$^{208}$Pb&$^{258}$Rf&184-202&14-33&1804&187.8&0.725&0.02-0.19&[44]&AD \\
$^{48}$Ca&$^{238}$U&$^{286}$Cn&185-215&26-56&1840&187.2&0.713&0.00025-0.125&[51]&MY \\
\label{table1}
\end{tabular}
\end{table} 
\end{center}

Using fissility as a typical scaling variable, I show a combined picture of all the data represented in Table 1 as Figure 6. There appears to be no easily discerned variation of P$_{CN}$ upon fissility in this plot.   Restricting attention to a limited region of excitation energy (E$^{*}$ = 40 -50 MeV) results in a clearer pattern as shown in Figure 7.  If we further note that the ``discordant" points in Figure 7 near  x$_{eff}$ =0.6 are cases where the projectile is the doubly magic $^{48}$Ca, then we might postulate, as others have done [42], that there are nuclear structure effects on P$_{CN}$.  Neglecting these nuclear structure effects leads to a straightforward dependence of P$_{CN}$ upon fissility shown by the lines in figure 7.  Clearly any estimates of P$_{CN}$ based upon the excitation energy and fissility dependences outlined herein are uncertain to at least an order of magnitude.

\begin{figure}[h]
\begin{minipage}{18pc}
\includegraphics[width=18pc]{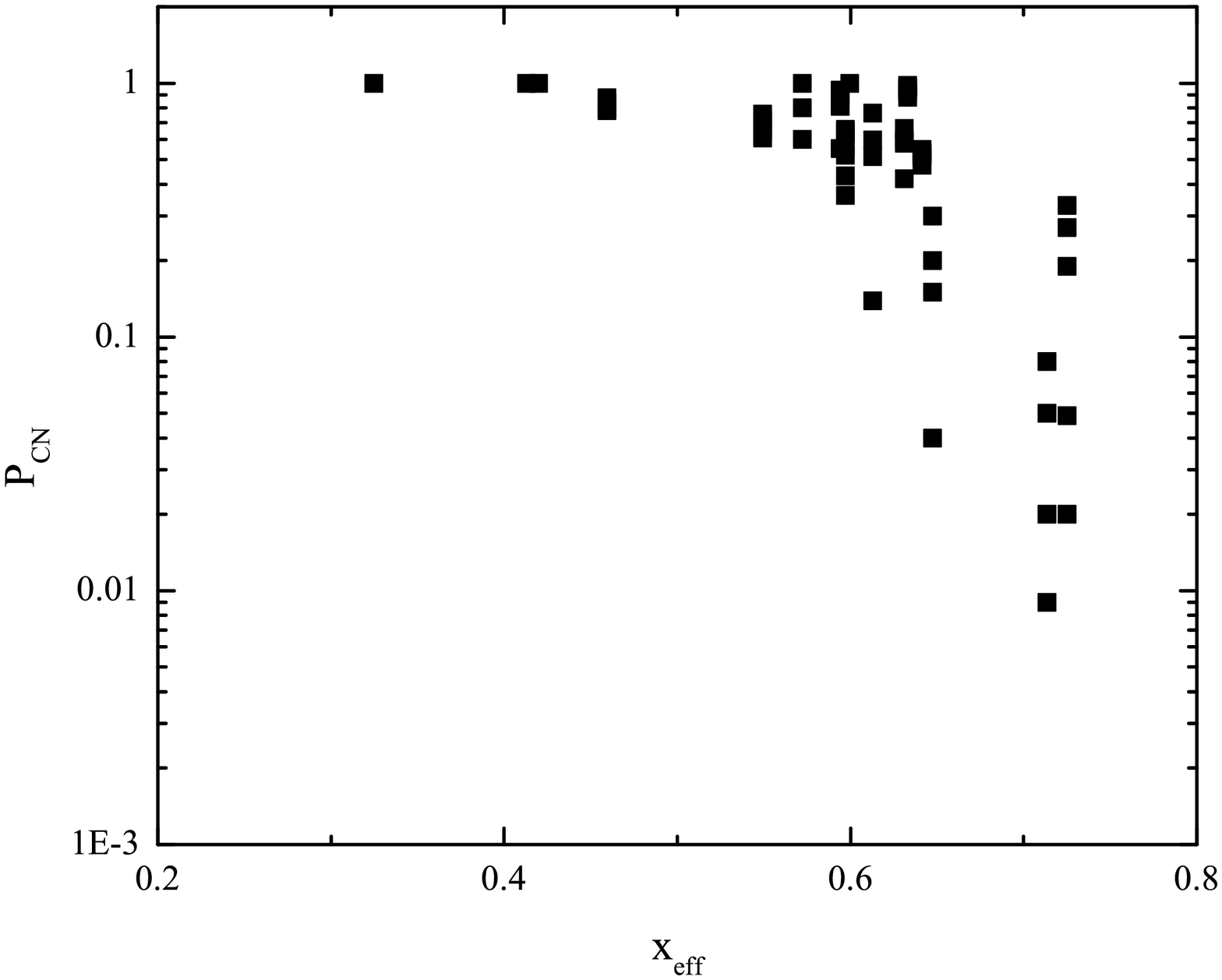}
\caption{\label{fig 6}Dependence of measured values of P$_{CN}$ upon fissility}
\end{minipage}\hspace{2pc}%
\begin{minipage}{18pc}
\includegraphics[width=18pc]{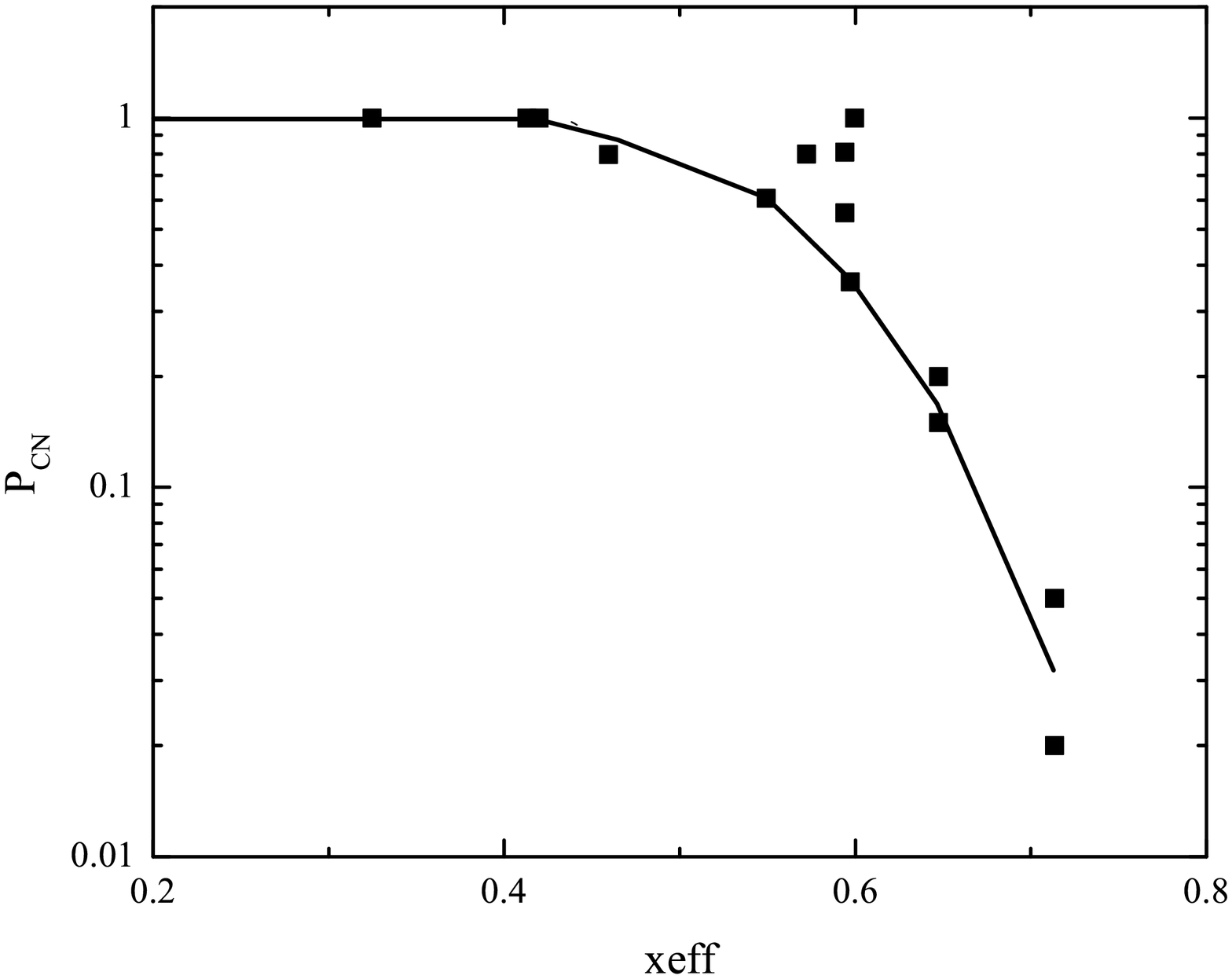}
\caption{\label{fig 7}Same as Figure 6 except values have been restricted to E* = 40-50 MeV}
\end{minipage} 
\end{figure}

\subsection{Predictions}

If we accept the formulation discussed above ( a coupled channels calculation of $\sigma$$_{capture}$, the classical treatment of W$_{sur}$ and the semi-empirical treatment of P$_{CN}$), we can make predictions as to the outcome of various complete fusion reactions.  These predictions are shown in Table 2.  Please remember these predictions are based upon the use of the global mass tables of [16] and they are uncertain to at least one order of magnitude.

\begin{center}
\begin{table}[h]
\centering
\caption{\label{predictions}Predicted empirical complete fusion cross sections for superheavy element production} 
\begin{tabular}{@{}l*{15}{l}}
\br
Reaction&$\sigma$$_{EVR}$(pb)\\
\mr
$^{249}$Bk($^{48}$Ca,3n)$^{294}$117&1\\
$^{249}$Bk($^{50}$Ti,4n)$^{295}$119&0.07\\
$^{248}$Cm($^{54}$Cr,4n)$^{302}$120&0.02\\
$^{244}$Pu($^{58}$Fe,4n)$^{302}$120&0.006\\
$^{238}$U($^{64}$Ni,3n)$^{302}$120&0.004\\
\br
\end{tabular}
\end{table}
\end{center}

\section{Alternate reactions}  
The predicted cross sections for the formation of new superheavy elements shown in Table 2 indicate that it will be challenging to pursue the synthesis of new heavy elements.  However, significant insights into the behavior of the heaviest elements may be possible by studying the more n-rich nuclei of nuclei of lower Z which have as yet not been synthesized.  I discuss a few of these synthetic pathways
\subsection{Damped collisions}
 Recently , there has
been a revival of interest in the use of damped collisions of massive nuclei
at near barrier energies to synthesize superheavy nuclei, particularly
those nuclei with large neutron excess, approaching the N=184 shell. In the
1980s [50] there were attempts to use the $^{238}$U + $^{238}$U and
the $^{238}$U + $^{248}$Cm reactions at above barrier energies to produce trans-target nuclides. While
there was evidence for the formation of neutron-rich isotopes of Fm and Md
at the 0.1 $\mu $b level, no higher actinides were found. The fundamental
problem was that the nuclei that were produced far above the target nucleus
were the result of events with high total kinetic energy loss, i.e., high
excitation energies and resulting poor survival probabilities. Very
recently, Zagrebaev and Greiner [51-58] using a new
model [59] for these collisions, have examined the older
experiments and some proposed new experiments ($^{232}$Th +$^{250}$Cf, $%
^{238}$U+$^{238}$U, and $^{238}$U +$^{248}$Cm). With their new model which
emphasizes the role of shell effects in damped collisions, they are able to
correctly describe the previously measured fragment angular, energy and
charge distributions from the $^{136}$Xe + $^{209}$Bi reaction and the
isotopic yields of Cf , Es, Fm and Md from the $^{238}$U + $^{248}$Cm
reaction. They predict that by a careful choice of beam energies and
projectile-target combinations, one might be able to produce n-rich isotopes
of element 112 in the $^{248}$Cm +$^{250}$Cf reaction. They suggest the
detection of $^{267,268}$Db and $^{272,271}$Bh (at the pb level) in the Th +
Cf or U + Cm reactions to verify these predictions. Such experiments are
very difficult because of the low cross sections, the lower intensities of
these massive projectile beams and the problems of detecting the reaction
products in an ocean of elastically scattered particles, etc.

However, in 2007, Zagrebaev and Greiner [63] outlined a simpler
test of their theoretical predictions. They applied the same model used to
study the U + Cm, Th + Cf and U + U collisions to the $^{160}$Gd + $^{186}$W
reaction. As an experimentalist, I really appreciate this suggestion of a surrogate reaction that allows one to check the theoretical predictions in a more accessible  system. 

In Figure [8] I show the results of an experimental study using radiochemical techniques of the $^{160}$Gd + $^{186}$W reaction.[64]  Both the measured and predicted mass distributions show the expected ``rabbit ears". i.e, a peak in the yields near the mass of the target and the projectile nuclei.  The measured distribution shows yields of what are probably fission fragments and intermediate mass fragments not predicted by the model.  Perhaps the most significant feature of the mass distribution is a peak in the mass distribution for trans target nuclei (A= 190-200)   A closer look at that peak is shown in Figures [9] and [10].  This trans target peak appears to be at Z=79 (Au), reminding one of the ``goldfinger" seen in studies of low energy deep inelastic scattering in the 1970s.  All this is consistent of the formation of a much heavier product that decays by fission and then particle emission to give rise to this trans target peak.  Zagrebaev and Greiner had actually predicted enhanced trans target yields in the Pb isotopes, which were searched for but not observed.  This result and the results of the TAMU group [65,66] for the 7.5 A MeV $^{197}$Au + $^{232}$Th reaction are encouraging for the effort to use these reactions to produce new n-rich heavy nuclei.

\begin{figure}[h]
\begin{minipage}{28pc}
\begin{center}
\includegraphics[width=28pc]{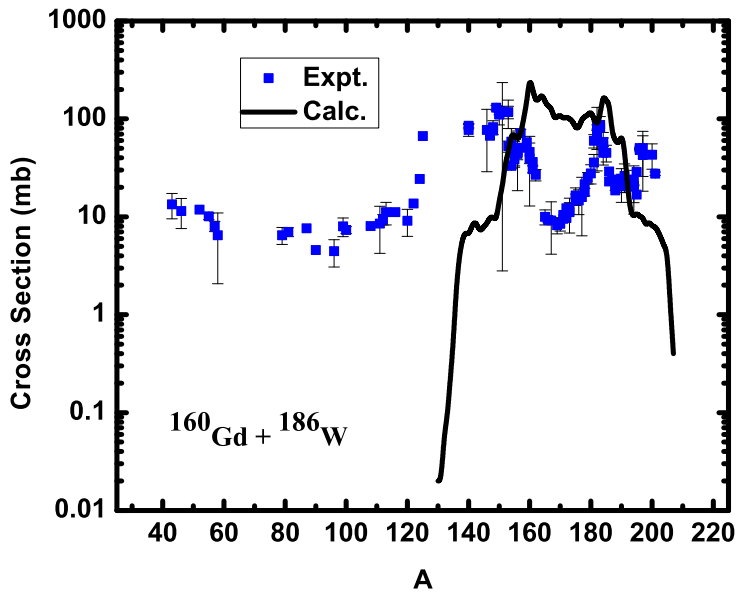}
\end{center}
\caption{\label{fig8}Comparison of measured [64] and predicted [63] values of the fragment mass distribution for the reaction of $^{160}$Gd with $^{186}$W.}
\end{minipage} 
\end{figure}

\begin{figure}[h]
\begin{minipage}{18pc}
\includegraphics[width=18pc]{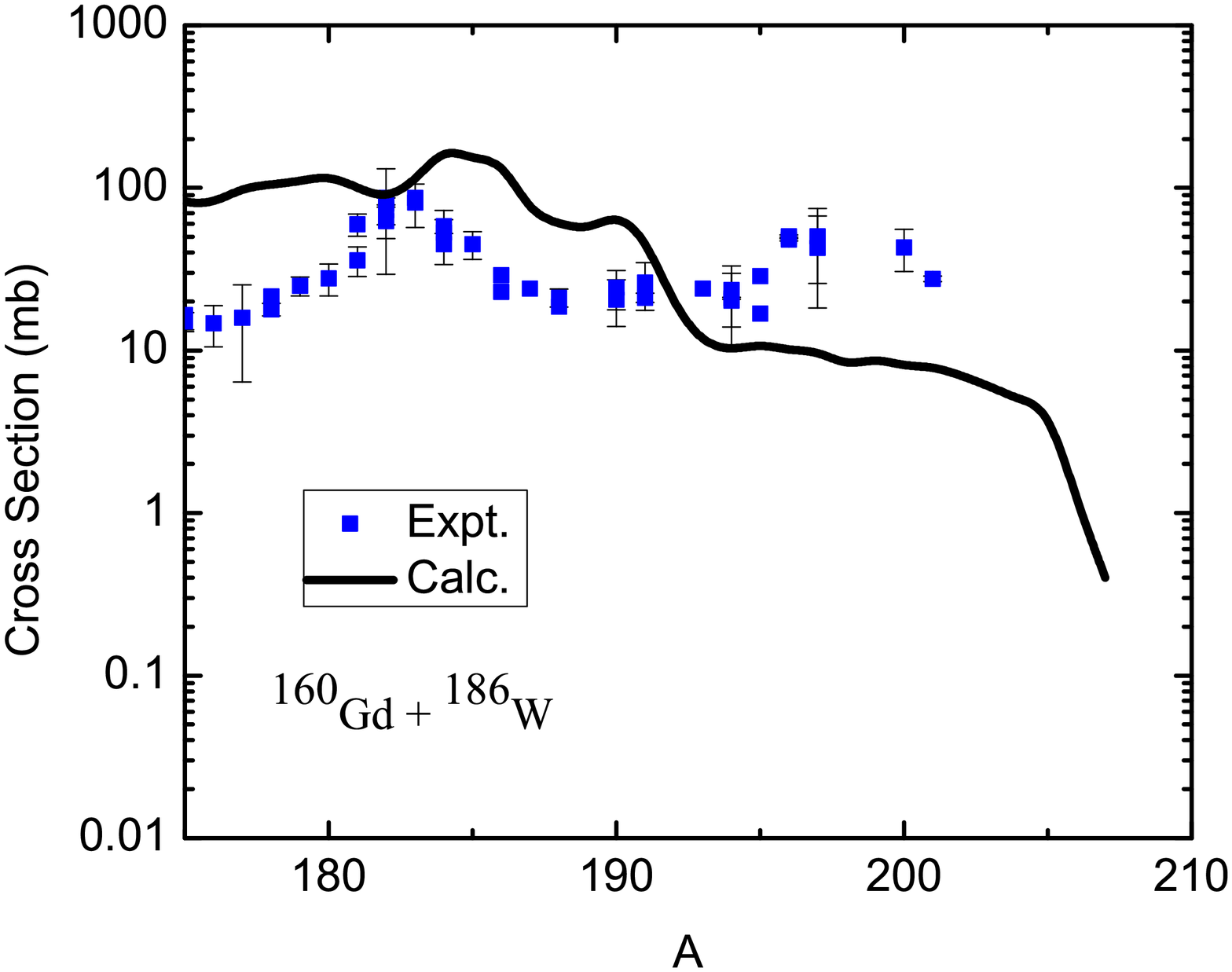}
\caption{\label{fig 1a}Comparison of measured [64] and predicted [63] values of the fragment mass distribution for the reaction of $^{160}$Gd with $^{186}$W..}
\end{minipage}\hspace{2pc}%
\begin{minipage}{18pc}
\includegraphics[width=18pc]{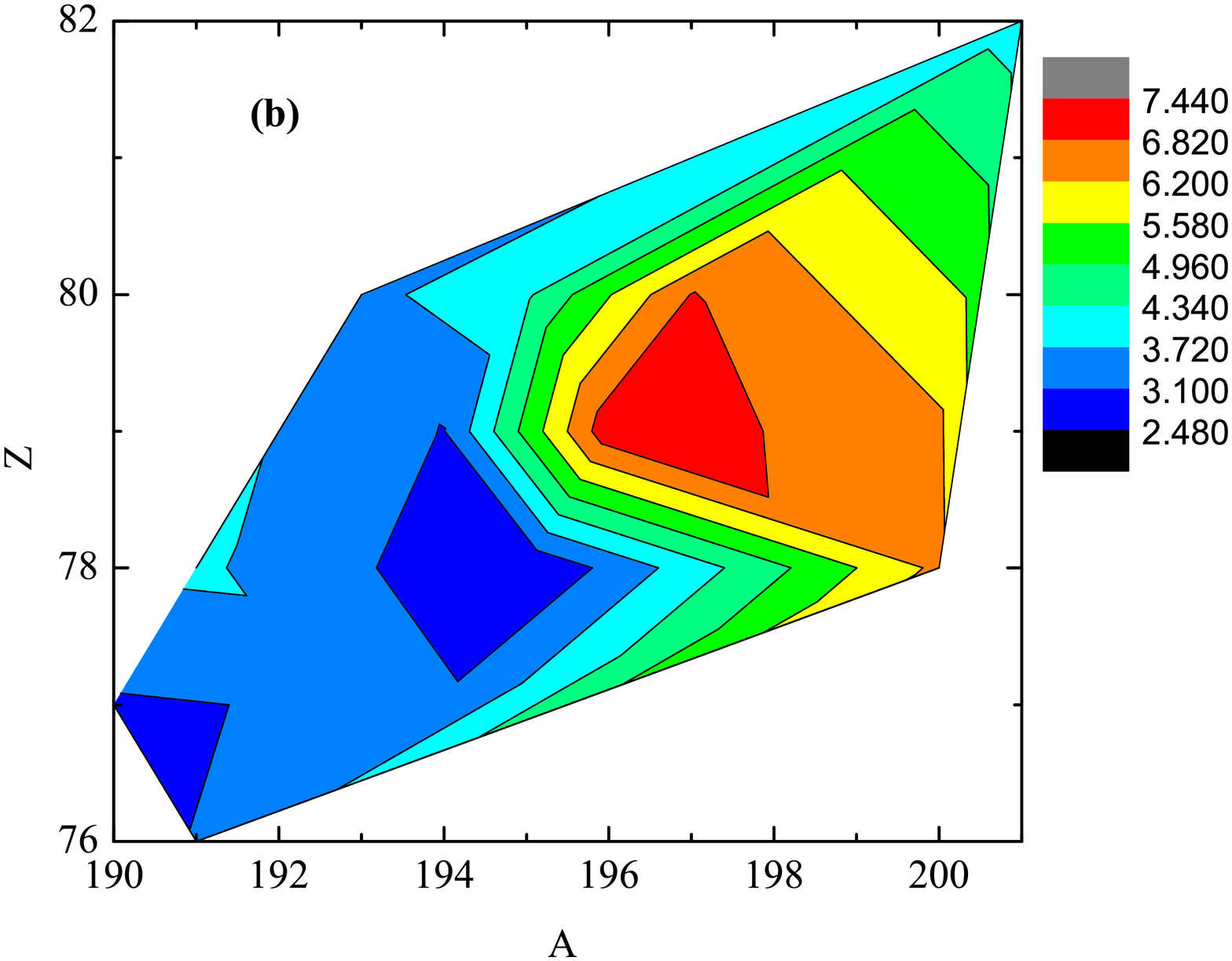}
\caption{\label{fig 1b}Contour plot of isotopic distribution of trans target nuclides.}
\end{minipage} 
\end{figure}

\section{Radioactive Beams}
The question can be posed as to why people want to use radioactive ion beams (RIBs) to produce new heavy nuclei.  The answers include : (a) The lower fusion barrier due to using  n-rich projectiles allows the synthesis reactions to take place using lower beam energies and hence lower excitation energies, i.e, leading to higher survival probabilities.  (b)  The formation of n-rich compound nuclei by itself leads to higher survival probabilities.  (c) The product nuclei have longer half-lives because the half-lives of the heaviest nuclei increase logarithmically with increasing neutron number and thus more detailed studies of the chemistry and atomic physics of the heaviest elements are enabled.  

We can apply what we know about the synthesis of the heaviest nuclei to the problem of making new heavy nuclei with radioactive beams.  The calculational model I employ [3] is simple and unsophisticated.  One takes the beam list for any radioactive beam facility (FRIB, SPIRAL2, ReA3, CARIBU, etc.[67-72]) and then considers every possible combination of a radioactive projectile with all ``stable" targets.  One varies the projectile energy and evaluates $\sigma$$_{capture}$, P$_{CN}$ and W$_{sur}$ to get $\sigma$$_{EVR}$.  From this, one uses reasonable assumptions about target thickness (0.5 mg/cm$^{2}$) and calculates the product yield in atoms/day.  

Some insight into the nature of the problem can be gained without doing any calculations but just looking at the intensities of typical radioactive beams from modern facilities.  In Figure [11] I show typical reported beam intensities for the re-accelerated rare gas beams available from several facilities.  (In this tabulation, i also include the predicted re-accelerated beams intensities for the defunct RIA project, a US facility that was too expensive to build).  I choose to compare the rare gas beams as the release times and subsequent losses for these beams should be minimal.  In each sub-panel of Figure [11] I show a horizontal line corresponding to a beam intensity of 1 particle micro ampere, a beam intensity that represents current stable beam facilities.  For the Ne beams one sees the radioactive beam intensities are at least 2-3 orders of magnitude less than the stable beam intensities and the intensities become far less as one goes further n-rich.  A similar situation occurs with the Ar beams except the radioactive beam intensities are even less.  For the Kr beams (which represent typical n-rich fission fragment beams) , the RIA concept was predicted to produce radioactive beams with intensities approaching those of stable beams.  The ISOL facility SPIRAL2 is projected to produce n-rich Kr beams that are about 2 orders of magnitude below stable beam intensities but the PF facilities such as FRIB and ReA3 have low intensities of these beams.  The $^{252}$Cf facility CARIBU is predicted to produce beam intensities of 10$^{4}$-10$^{5}$ particle/sec of the n-rich Kr isotopes.

\begin{figure}[h]
\begin{minipage}{40pc}
\begin{center}
\includegraphics[width=40pc]{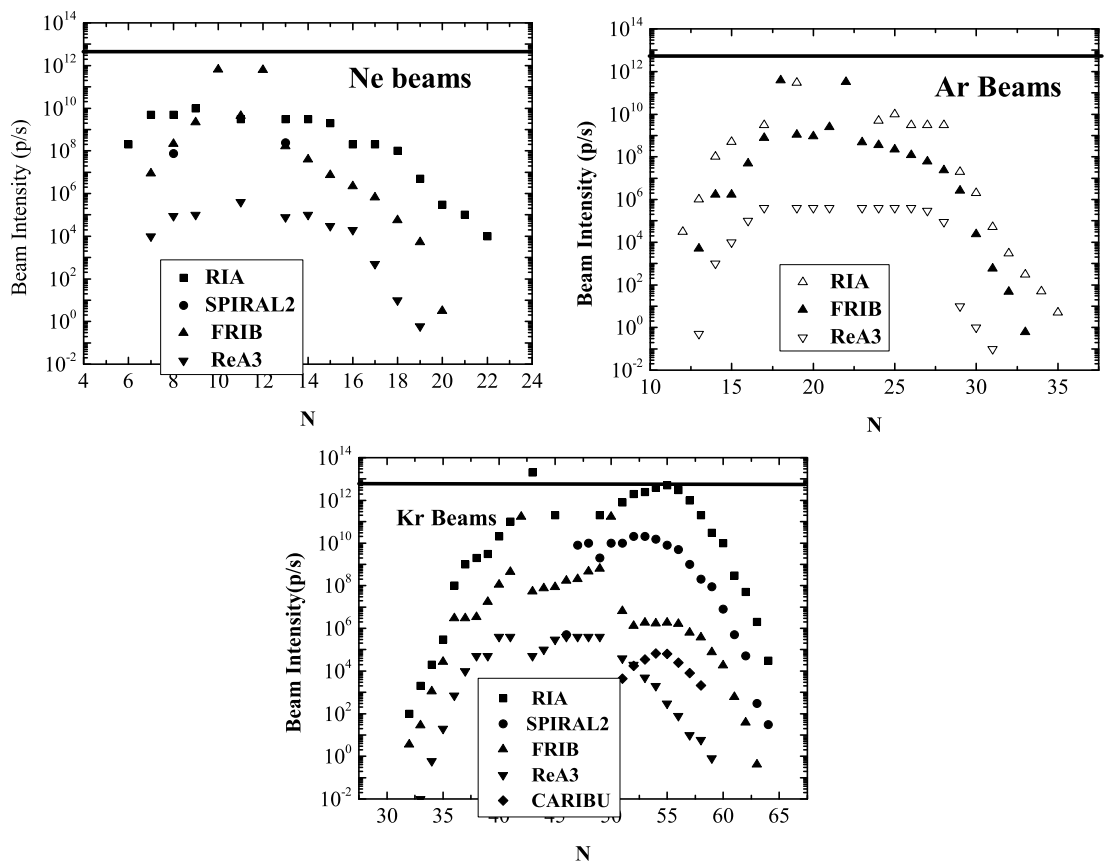}
\end{center}
\caption{\label{fig11}Rare gas beam intensities from several radioactive beam facilities}
\end{minipage} 
\end{figure}

An important {\it caveat} about these beam intensities should be noted.  The quoted beam intensities represent the predicted intensities of re-accelerated beams.  This choice represents a conservative cautious approach to the possible beam intensities.  Some have advocated a different more aggressive approach in which one takes, for PF facilities, such as FRIB, the ``fast" beam intensities which are typically 2-3 orders of magnitude larger than the re-accelerated beam intensities and assumes that one will be able to deliver these beams at Coulomb barrier energies with 10$\%$ efficiency.  A related issue is that of ``targeted beams", i.e., special beams  of such importance that campaigns could be created to produce these beams by special developments.  For example, several studies of heavy element synthesis reactions involve the use of $^{48}$Ca beams of intensities of the order of 1 particle micro ampere (with projected future developments of beams of 10 particle micro amperes.)  One could conceive of efforts to produce n-rich K beams by fragmentation or transfer reactions involving the use of $^{48}$Ca beams.

In Figures 12 and 13 I show a comparison of the best stable beam reaction production rate vs. the best radioactive beam production rate for cold and hot fusion reactions.  The radioactive beam production rates are 3 orders of magnitude below the stable beam production rates.  { \bf Radioactive beams are not a pathway to new superheavy elements.}  Does that mean that radioactive beams are worthless when it comes to making new heavy nuclei?  No, radioactive beams are useful tools for producing new n-rich isotopes of elements 104-107.  In Table 3  I show a list of new n-rich isotopes of elements 104-107 that can be made at rates greater than 5 atoms/day and the reactions that produce them.

\begin{figure}[h]
\begin{minipage}{20pc}
\includegraphics[width=20pc]{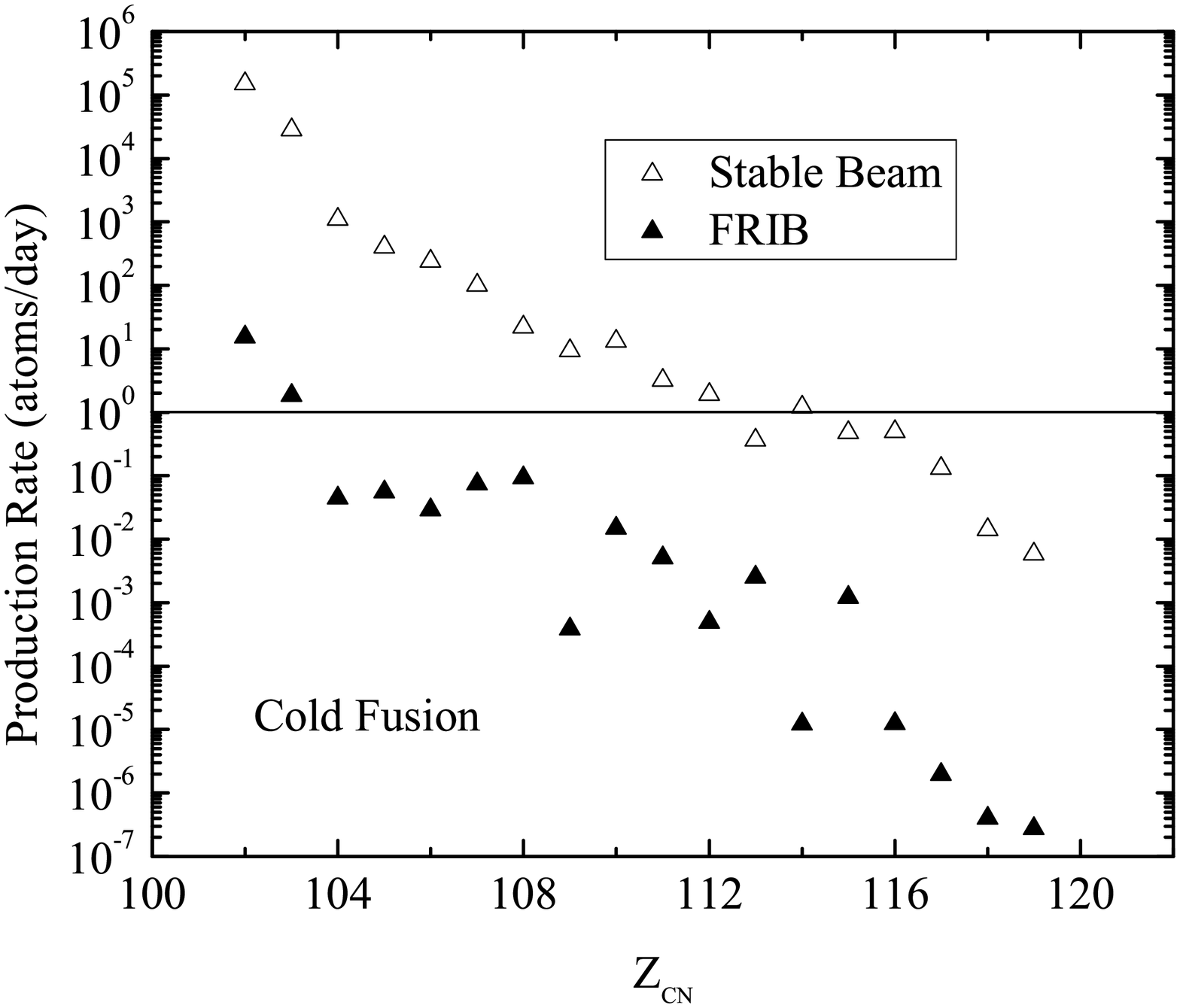}
\caption{\label{fig 12}Predicted heavy element production rates using stable and radioactive beams for cold fusion reactions}
\end{minipage}\hspace{2pc}%
\begin{minipage}{20pc}
\includegraphics[width=20pc]{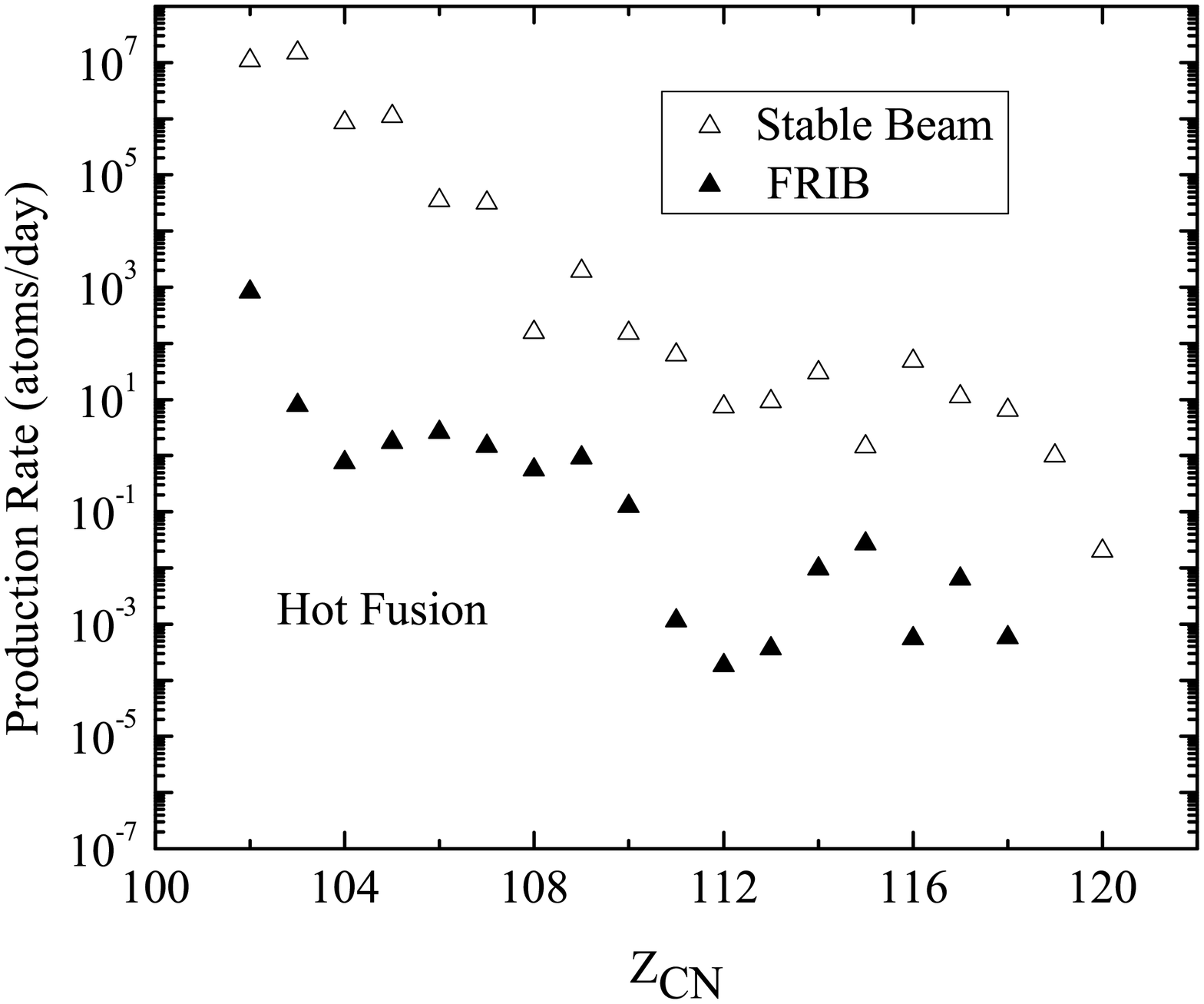}
\caption{\label{fig 13}Predicted heavy element production rates using stable and radioactive beams for hot fusion reactions}
\end{minipage} 
\end{figure}

\begin{center}
\begin{table}[h]
\centering
\caption{\label{fiveatoms}Reactions predicted to form 5 or more atoms per day of new n-rich nuclei} 
\begin{tabular}{@{}l*{15}{l}}
\br
Nucleus&Reaction\\
\mr
$^{264}$Rf&$^{252}$Cf($^{16}$C,4n)\\
$^{265}$Db&$^{249}$Bk($^{20}$O,4n)\\
$^{268}$Sg&$^{252}$Cf($^{20}$O,4n)\\
$^{267}$Bh &$^{252}$Cf($^{21}$F,4n)\\

\br
\end{tabular}
\end{table}
\end{center}

One might pose the question as to which radioactive beams are projected to be the most useful in synthesizing these n-rich nuclei.  The answer to this question is the light beams such as O, Ne, Mg, etc. because of their high intensities.  In Table 4 I show the reactions and rates for the production of new n-rich isotopes of Sg, which all involve these light nuclei.

\begin{table}[h]
\caption{\label{ex}Typical reactions that form new n-rich isotopes of Sg.}
\begin{center}
\begin{tabular}{llll}
\br
Reactants&Products &FRIB beam intensity (p/s)&Production Rate (atoms/day)\\
\mr
$^{26}$Ne + $^{248}$Cm&$^{271}$Sg + 4n&2.2 x 10$^{6}$&0.004\\
$^{30}$Mg + $^{244}$Pu&$^{270}$Sg + 4n&7.1 x 10$^{6}$&1\\
$^{29}$Mg + $^{244}$Pu&$^{269}$Sg + 4n&3.6 x 10$^{7}$&0.2\\
$^{20}$O + $^{252}$Cf&$^{268}$Sg + 4n&1.5 x 10$^{8}$&5\\
$^{23}$Ne + $^{248}$Cm&$^{267}$Sg + 4n&1.6 x 10$^{8}$&1\\
\br
\end{tabular}
\end{center}
\end{table}

I conclude that:  (a) RNBs offer unique opportunities to explore the physics and chemistry of n-rich heavy nuclei in the short and long term.  (b)RNBs are not a path to new chemical elements (c)  RNB research can help us to understand the isospin dependence of fundamental quantities in heavy element science.

\section{Conclusions} I conclude that:
(a) New directions in synthesizing heavy nuclei can be pursued to make n-rich heavy nuclei with transfer reactions and reactions with radioactive beams.
(b) There is work to be done to understand the physics of the fusion reactions used to date.
(c) Heavy element synthesis studies remain a laboratory for studying nuclei, their structure and reactions at the limits of stability.
\section{Acknowledgements}
This work was supported in part by the Office of High Energy and Nuclear Physics, Nuclear Physics Division, US Department of Energy, under Grant No. DE-FG06-97ER41026
\section{References}

\smallskip

\end{document}